\documentclass[runningheads]{llncs}
\usepackage{acronym}
\usepackage{graphicx}
\usepackage{epstopdf}
\usepackage{subfigure}
\usepackage{booktabs}
\usepackage{listings}
\usepackage{amsmath}

\acrodef{BML}{Biham-Middleton-Levine}
\acrodef{CA}{Cellular Automaton}
\acrodef{GPU}{Graphics Processing Unit}
\acrodef{SIMD}{Single Instruction Multiple Data}

\begin{document}
\title{Parallel Implementations of Cellular Automata for Traffic Models}
\author{Moreno Marzolla\orcidID{0000-0002-2151-5287}}
\authorrunning{M. Marzolla}
\institute{Department of Computer Science and Engineering\\
University of Bologna, Italy\\
\email{moreno.marzolla@unibo.it}}

\maketitle

\begin{abstract} 
The Biham-Middleton-Levine (BML) traffic model is a simple
two-dimensional, discrete Cellular Automaton~(CA) that has been used
to study self-organization and phase transitions arising in traffic
flows. From the computational point of view, the~BML model exhibits
the usual features of discrete~CA, where the state of the automaton
are updated according to simple rules that depend on the state of each
cell and its neighbors. In this paper we study the impact of various
optimizations for speeding up~CA computations by using the~BML model
as a case study. In particular, we describe and analyze the impact of
several parallel implementations that rely on~CPU features, such as
multiple cores or~SIMD instructions, and on~GPUs. Experimental
evaluation provides quantitative measures of the payoff of each
technique in terms of speedup with respect to a plain serial
implementation. Our findings show that the performance gap between~CPU
and~GPU implementations of the~BML traffic model can be reduced by
clever exploitation of all CPU features.

\end{abstract}

\section{Introduction}

Cellular Automata~(CA)\acused{CA} are a simple computational model of
many natural phenomena, such as virus infections in biological
systems, turbulence in fluids, chemical reactions~\cite{Gutowitz1991}
and traffic flows~\cite{Maerivoet:2005}. In its simplest form, a
discrete~\ac{CA} consists of a finite lattice of cells (domain), where
each cell can be in any of a finite number of states. The cells evolve
synchronously at discrete points in time; the new value of a cell
depends on its previous value and on the values of its neighbors
according to some fixed rule.

Simulating the evolution of~\ac{CA} models can be computationally
challenging, especially for large domains and/or complex update rules.
However, many~\ac{CA} models belong to the class of
\emph{embarrassingly parallel computations}, meaning that new states
can be computed in parallel if multiple execution units are available.
This is actually the case: virtually every desktop- or server-class
processor on the market today provides advanced parallel capabilities,
such as multiple execution cores and~\ac{SIMD} instructions. Moreover,
programmable~\acp{GPU} are ubiquitous and affordable, and are
particularly suited for this kind of applications since they provide a
large number of execution units that can operate in parallel.

Unfortunately, exploiting the computational power of modern
architectures requires programming techniques and specialized
knowledge that are not as diffuse as they should be. Additionally,
there is considerable misunderstanding about which programming
technique and/or parallel architecture is the most effective in any
given situation. This results in many exaggerated claims that later
proved to be unsubstantiated~\cite{Lee2010}.

In this paper we study the impact of various optimizations for
speeding up~\ac{CA} computations by using the~\ac{BML} model as a case
study. We focus on two common computing architectures:
(\emph{i})~multicore CPUs, i.e., processors with multiple independent
execution units, and (\emph{ii})~general-purpose~\acp{GPU} that
include hundreds of simple execution units that can be programmed for
any kind of computation. Starting with an unoptimized~CPU
implementation of the~\ac{BML} model, we develop incremental
refinement that incorporate more advanced features: shared-memory
programming, \ac{SIMD} instructions, and a full~\ac{GPU}
implementation. We compare experimentally the various versions on two
machines to analyze the impact of each optimization. Our findings show
that, for the~\ac{BML} model, CPUs can be extremely effective if all
their features are correctly exploited. We believe that the findings
reported in this paper can be useful for improving the simulation of
other, more realistic, traffic models based on~Cellular Automata.

This paper is organized as follows: in Section~\ref{sec:bml-model} we
briefly describe the main features of the~\ac{BML} model. In the next
sections we start with a simple serial implementation of the model
(Section~\ref{sec:serial}), that is later improved to take advantage
of shared-memory parallelism (Section~\ref{sec:openmp}), of
the~\acl{SIMD} programming paradigm (Section~\ref{sec:simd}), and
of~\acp{GPU} (Section~\ref{sec:cuda}). In
Section~\ref{sec:performance-evaluation} we compare the performance of
all the implementations above on two typical mid-range
machines. Finally, the conclusions of this work are reported in
Section~\ref{sec:conclusions}.

\section{The Biham-Middleton-Levine traffic model}\label{sec:bml-model}

The~\ac{BML} model~\cite{Biham:1992} is a simple~\ac{CA} that
describes traffic flows in two dimensions. In its simplest form, the
model consists of a periodic square lattice of $N \times N$
cells. Each cell can be either empty, or occupied by a vehicle moving
from top to bottom (\texttt{TB}) or from left to right
(\texttt{LR}). The model evolves by alternating horizontal and
vertical phases. During a horizontal phase, all \texttt{LR} vehicles
move one cell right, provided that the destination cell is empty.
Similarly, during a vertical phase, all \texttt{TB} vehicles move one
cell down the grid. A vehicle exiting the grid from one side reappears
on the opposite side, therefore realizing periodic boundary
conditions.

\begin{figure}
  \centering
  \subfigure[\label{fig:bml:free-flowing}]{\includegraphics[width=.3\textwidth]{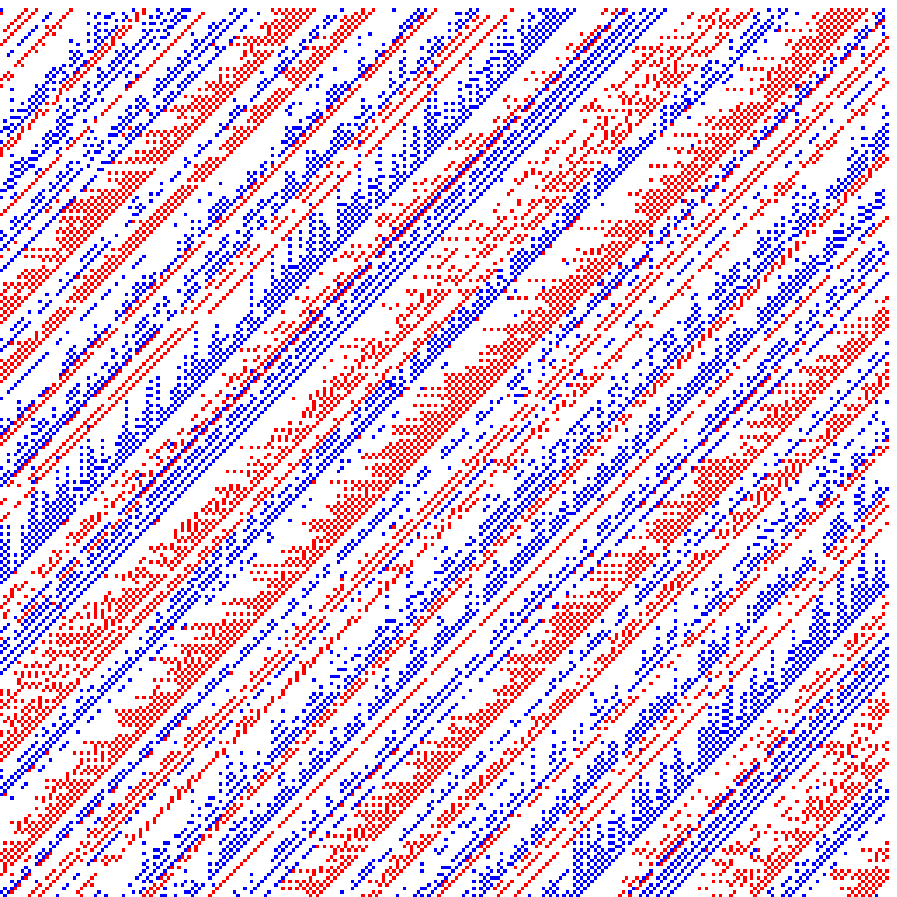}}\quad%
  \subfigure[\label{fig:bml:intermediate}]{\includegraphics[width=.3\textwidth]{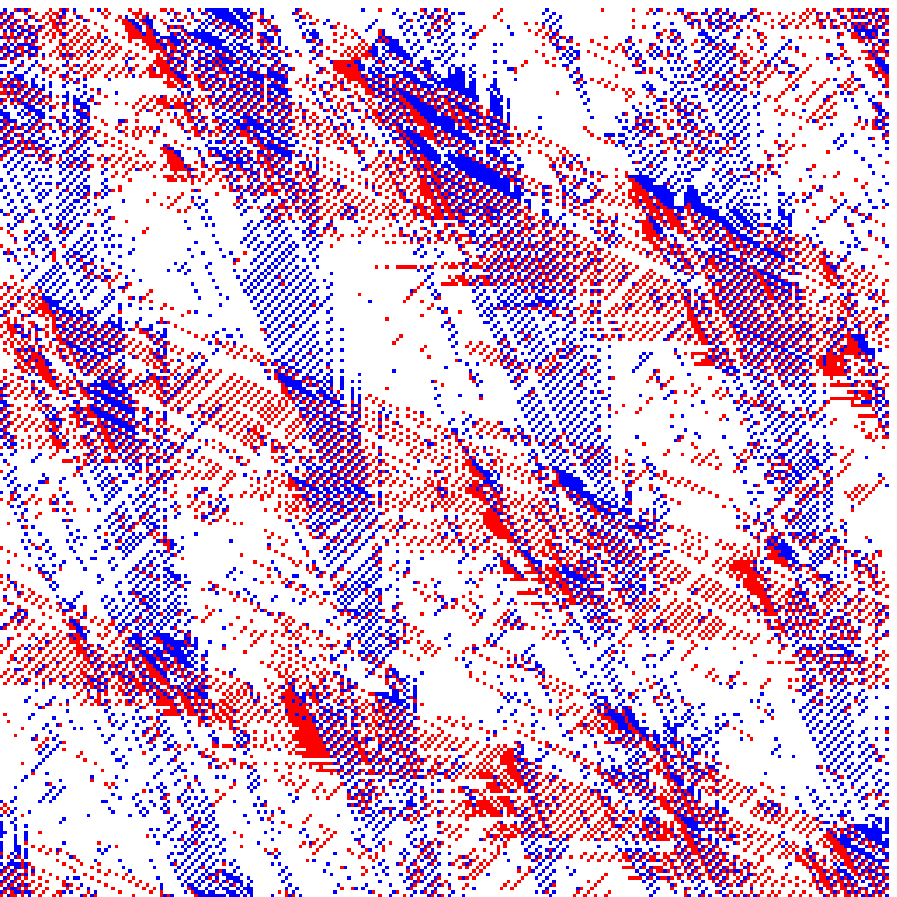}}\quad%
  \subfigure[\label{fig:bml:jammed}]{\includegraphics[width=.3\textwidth]{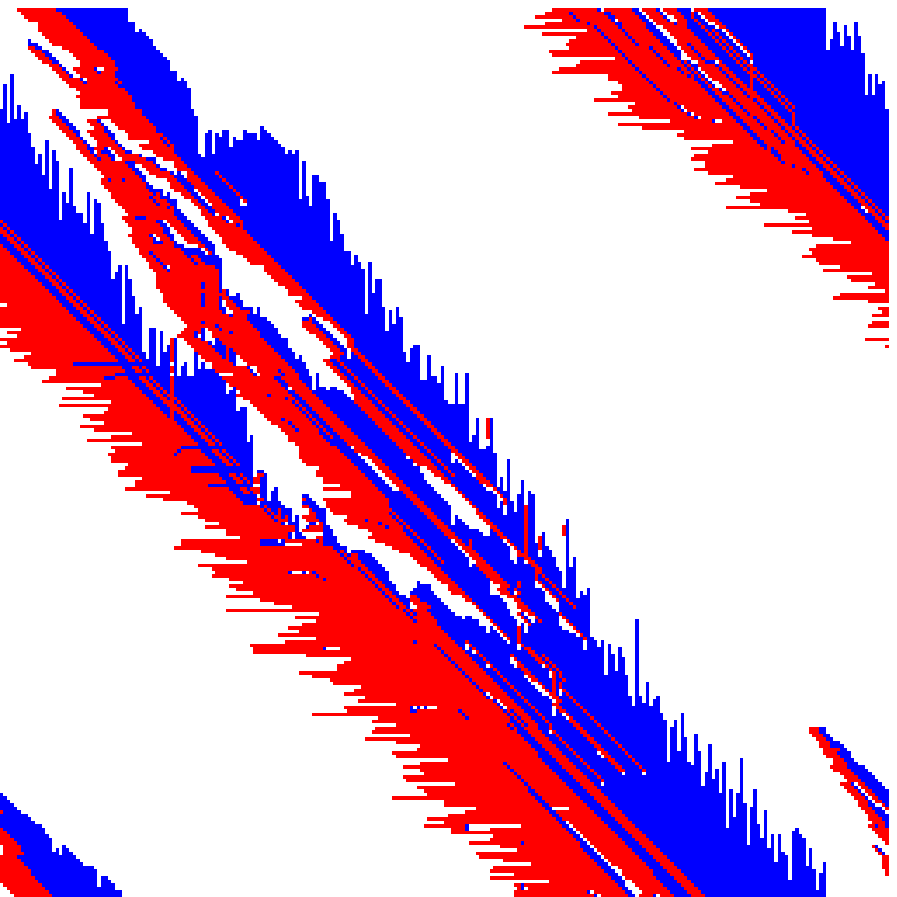}}
  \caption{BML model on a $256 \times 256$ lattice after~$4096$
    steps. (a)~Free-flowing phase, $\rho = 0.25$; (b)~Intermediate
    phase, $\rho = 0.32$; (c)~Globally jammed phase, $\rho =
    0.38$. Red dots represent \texttt{LR} vehicles, while blue dots
    represent \texttt{TB} vehicles.}\label{fig:bml}
\end{figure}

Despite its simplicity, the~\ac{BML} model undergoes a phase
transition when the density~$\rho$ of vehicles exceeds a critical
value that depends on the grid
size~$N$~\cite{Biham:1992,DSouza:2005}. When~$\rho$ is below the
critical threshold, the system stabilizes in a \emph{free-flowing
  state} where vehicles arrange themselves in a non-interfering
pattern to achieve maximum average speed. If the density is just above
the critical threshold, a global jam eventually develops and no
further movement is possible.

Figure~\ref{fig:bml} shows three configurations of the~\ac{BML} model
after~$4096$ steps (each step includes a horizontal and vertical
phase) on a grid of size $256 \times 256$ for different values of the
vehicle density~$\rho$. There are approximately $\rho N/2$ vehicles of
each type that are initially placed randomly on the
grid. Figure~\ref{fig:bml:free-flowing} shows the free-flowing state
that arises when $\rho = 0.25$. Increasing the vehicle density $\rho =
0.32$ Figure~\ref{fig:bml:intermediate} shows the intermediate phase
that can be observed if the value of~$\rho$ is increased, but is still
below the critical threshold. When the density~$\rho$ reaches the
critical threshold, all vehicles are eventually stuck in a giant
traffic jam as shown in Figure~\ref{fig:bml:jammed}, and the average
speed drops to zero.
In fact, the behavior of the~\ac{BML} model is more complex: the
free-flowing and jammed states might coexist, i.e., have non-zero
probability to occur, when~$\rho$ lies within some interval around
the critical point~\cite{DSouza:2005}.

The following rule can be used during a horizontal phase to compute
the new state \verb+center'+ of a cell, given its current state
\verb+center+ and the current state of its \verb+left+ and
\verb+right+ neighbors:

\begin{align*}
\texttt{center'} &= \begin{cases}
\texttt{LR} & \text{if}\ \texttt{left} = \texttt{LR} \wedge \texttt{center} = \texttt{EMPTY} \\
\texttt{EMPTY} & \text{if}\ \texttt{center} = \texttt{LR} \wedge \texttt{right} = \texttt{EMPTY} \\
\texttt{center} & \text{otherwise}
\end{cases}
\end{align*}

Similarly, the following rule can be used to compute the new state of
a cell during a vertical phase, given its current state and the state
of its neighbors locate at the \verb+top+ and \verb+bottom+:

\begin{align*}
\texttt{center'} &= \begin{cases}
\texttt{TB} & \text{if}\ \texttt{top} = \texttt{TB} \wedge \texttt{center} = \texttt{EMPTY} \\
\texttt{EMPTY} & \text{if}\ \texttt{center} = \texttt{TB} \wedge \texttt{bottom} = \texttt{EMPTY} \\
\texttt{center} & \text{otherwise}
\end{cases}
\end{align*}

Two variations are described in the original paper~\cite{Biham:1992},
called Model~II and Model~III. Model~II allows all vehicles to advance
one step in the same phase; if a~\texttt{LR} and~\texttt{TB} vehicle
try to occupy the same empty cell, one of them is randomly selected to
move while the other stands still. Model~III allows the same cell to
contain both a~\texttt{LR} and a~\texttt{TB} vehicle. Both these
models exhibit the same behavior of Model~I.

The original~\ac{BML} model has been extended in several
directions. C\'ampora et al.~\cite{Campora:2010} studied the behavior
of the automaton on a square lattice embedded in a Klein bottle. Ding
et al.~\cite{Ding:2011} analyzed a different update rule where a
randomly chosen vehicle is allowed to move at each step. Freund and
P{\"o}schel~\cite{Freund:1995} proposed a more general automaton where
vehicles can move in all directions and intersections are modeled more
realistically. Other extensions of the~\ac{BML} model
include~\cite{Chowdhury:2000}: asymmetric distribution of vehicles where
the number of \texttt{TB} and \texttt{LR} vehicles is not the same;
different speeds for \texttt{TB} and \texttt{LR} vehicles; two-level
crossing to simulate the presence of overpasses and underpasses;
permanent or transient road blocks, e.g., caused by traffic accidents.

\section{Serial implementation}\label{sec:serial}

The~\ac{BML} model is a \emph{synchronous}~\ac{CA}, since it requires
that all cells are updated at the same time. To achieve this it is
necessary to use two grids, say \verb+cur+ and \verb+next+, holding
the current and next~\ac{CA} configuration, respectively. Cell states
are read from \verb+cur+\, and new states are written to
\verb+next+. When all new states have been computed, \verb+cur+ and
\verb+next+ are exchanged.

Since the~C language lays out~2D arrays row-wise in memory, it is
easier to treat \verb+cur+ and \verb+next+ as 1D arrays, and use a
function (or macro) \verb+IDX(i,j)+ to compute the mapping of the
coordinates $(i, j)$ to a linear index. Therefore, we write
\verb+cur[IDX(i,j)]+ to denote the cell at coordinates $(i, j)$ of
\verb+cur+. In case of a $N \times N$ grid, the function
\verb+IDX(i,j)+ will return $(i \times N + j)$.

Since the~\ac{BML} model is a three-state~\ac{CA}, two bits would be
sufficient to encode each cell. However, to simplify memory accesses
we use one byte per cell. The following code defines all necessary
data types, and shows how the horizontal phase can be realized (the
vertical phase is very similar, and is therefore omitted).

\begin{lstlisting}
typedef unsigned char cell_t;
enum {EMPTY = 0, LR, TB};
cell_t *cur, *next;

void horizontal_step(cell_t *cur, cell_t *next, int N) {
  int i, j;
  for (i=0; i<N; i++) {
    for (j=0; j<N; j++) {
      const cell_t left = cur[IDX(i,(j-1+N)%N)];
      const cell_T center = cur[IDX(i,j)];
      const cell_t right = cur[IDX(i,(j+1)%N)];
      cell_t *out = &next[IDX(i,j)];
      *out = (left == LR && center == EMPTY ? LR :
              (center == LR && right == EMPTY ? EMPTY :
               center));
    }
  }
}

void vertical_step(cell_t *cur, cell_t *next, int N) { ... }
\end{lstlisting}

\begin{figure}[t]
  \centering\includegraphics[scale=.7]{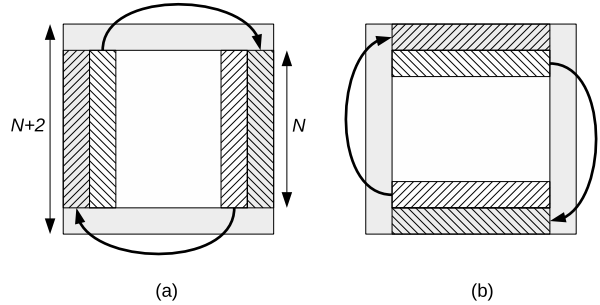}
  \caption{An~$N \times N$ domain (white) augmented with ghost cells
    (gray). The corners of the extended domain are ignored, since they
    are not used by the~BML update rules.}\label{fig:ghost}
\end{figure}

A direct implementation of the~\ac{BML} morel uses grids of $N \times
N$ elements. However, care must be taken when accessing the neighbors
of cell $(i, j)$ to avoid out-of-bound accesses. A common optimization
is to surround the domain with additional rows and columns, called
\emph{ghost cells}~\cite{Kjolstad:2010}. The domain becomes a grid of
size $(N+2) \times (N+2)$, where the true domain consists of the cells
$(i, j)$ for all $1 \leq i \leq N+1$, $1 \leq j \leq N+1$, while those
on the border contain a copy of the cells at the opposite side (see
Figure~\ref{fig:ghost}).

The top and bottom ghost rows must be filled before each vertical
update phase (Fig.~\ref{fig:ghost}~(a)), while the ghost columns on
the left and right must be filled before each horizontal phase
(Fig.~\ref{fig:ghost}~(b)). Ghost cells simplify the indexing of
neighbors and provide a significant speedup as will be shown in
Section~\ref{sec:performance-evaluation}.

\section{OpenMP implementation}\label{sec:openmp}

OpenMP~\cite{OpenMP} is an open standard that supports parallel
programming on shared-memory architectures; bindings for the C, C++
and FORTRAN languages are available. OpenMP allows the programmer to
annotate portions of the code as parallel regions; the compiler
generates the appropriate code to dispatch those regions to the
processor cores.

In the~C and~C++ languages, OpenMP annotations are specified using
\verb+#pragma+ preprocessor directives. One such directives is
\verb+#pragma omp parallel for+, that can be used to automatically
distribute the iterations of a ``for'' loop to multiple cores,
provided that the iterations are independent (this requirement must be
verified by the programmer). The update loop(s) of the~\ac{BML} model
can then be parallelized very easily as follows (we are assuming the
presence of ghost cells, so that the indexed $i$ and $j$ assume the
values $1, \ldots, N$).

\begin{lstlisting}
  ...
  #pragma omp parallel for
  for (i=1; i<N-1; i++) {
    for (j=1; j<N-1; j++) {
      /* update cell (i,j) */
    }
  }
  ...
\end{lstlisting}

\section{SIMD implementation}\label{sec:simd}

Modern processors provide~\ac{SIMD} instructions that can apply the
same mathematical or logical operation to multiple data items stored
in a register. A~\ac{SIMD} register is a small vector of fixed length
(usually, 128 or~256 bits). Depending on the processor capabilities, a
128-bit wide \ac{SIMD} register might contain, e.g., two 64-bit
doubles, four 32-bit floats, four 32-bit integers, and so on.

Some applications can greatly benefit from \ac{SIMD}
instructions. However, there are some limitations:
(\emph{i})~\ac{SIMD} instructions are processor-specific, and hence
not portable. (\emph{ii})~Automatic generation of~\ac{SIMD}
instructions from scalar code is beyond the capabilities of most
compilers and requires manual intervention from the
programmer. (\emph{iii})~\ac{SIMD} instructions might impose
constraints on how data is laid out in memory, e.g., by forcing
specific alignments for memory loads and stores.

The~SSE2 instruction set of Intel processors provides instructions
that can operate on~16 chars packed into a 128-bit~\ac{SIMD}
register. This allows us to compute the new states of~16 adjacent
cells at the time.  The~\ac{SIMD} version of the~\ac{BML} model has
been realized using \emph{vector data types} provided by the~GNU C
Compiler (GCC). Vector data types are a proprietary extension of~GCC
that allow users to use \ac{SIMD} vectors as if they were scalars: the
compiler emits the appropriate instructions to apply the desired
arithmetic or logical operator to all the elements of the vector.

The conditional branches required to compute the next state of each
cell are difficult to vectorize. The reason is that a branch may cause
a different execution path to be taken for different elements of
a~\ac{SIMD} register, which contrasts with the~\ac{SIMD} paradigm that
requires that the same sequence of operations is applied to all data
items. To overcome this problem we need to compute the new states
using a technique called \emph{selection and masking}, that makes use
of bit-wise operations only. The idea is to replace a statement like
\verb+a = (C ? x : y)+, where \verb+C+ is~$0$ or~$-1$
(\verb+0xffffffff+ in two's complement, hexadecimal notation), with
the functionally equivalent statement \verb+a = (C & x) | (~C & y)+,
where the conditional branch no longer appears.

The code below defines a vector datatype \verb+v16i+ of~16 chars and
uses it to compute the new state \verb+out+ of~16 adjacent cells at
the time.

\begin{lstlisting}
typedef char v16i __attribute__((vector_size(16)));
void horizontal_step(cell_t *cur, cell_t *next, int N) {
  int i, j;
  for (i=1; i<N+1; i++) {
    for (j=1; j<(N+1)-15; j += 16) {
      const v16i left = __builtin_ia32_loaddqu((char*)&cur[IDX(i,j-1)]);
      const v16i center = __builtin_ia32_loaddqu((char*)&cur[IDX(i,j)]);
      const v16i right = __builtin_ia32_loaddqu((char*)&cur[IDX(i,j+1)]);
      const v16i mask_lr = ((left == LR) & (center == EMPTY));
      const v16i mask_empty = ((center == LR) & (right == EMPTY));
      const v16i mask_center = ~(mask_lr | mask_empty);
      const v16i out = ((mask_lr & LR) | (mask_empty & EMPTY) | \
                        (mask_center & center));
      __builtin_ia32_storedqu((char*)&next[IDX(i,j)], out);
    }
  }
}
\end{lstlisting}

\verb+left+, \verb+center+ and \verb+right+ can be thought as C arrays
of length 16 holding the values of the left, center, and right
neighbors of~16 adjacent cells. Their contents are fetched from memory
using the \verb+__builtin_ia32_loaddqu+ (Load Double Quadword
Unaligned) intrinsic (again we are assuming that the domain is
extended with ghost cells).

The \verb+mask_lr+, \verb+mask_empty+ and \verb+mask_center+ vector
variables contain the value $-1$ in the positions where \verb+LR+,
\verb+EMPTY+ and \verb+center+ should be stored, respectively. Note
that the compiler automatically converts scalars to vectors, to handle
mixed comparisons like \verb+(left == LR)+. Conveniently, SSE2
comparison instructions generate the value $-1$ (instead of $1$) for
\emph{true}. The new states \verb+out+ of the~16 cells can then be
computed using bit-wise operators that the compiler translates into a
sequence of~\ac{SIMD} instructions. The result is stored back in
memory using the \verb+__builtin_ia32_storedqu+ (Store Double Quadword
Unaligned) intrinsic.

\section{CUDA implementation}\label{sec:cuda}

A modern~\ac{GPU} contains a large number of programmable processing
cores that can be used for general-purpose computing. The first widely
used framework for~\ac{GPU} programming has been the~CUDA toolkit by
NVidia corporation. A~CUDA program consists of a part that runs on
the~CPU and one that runs on the~\ac{GPU}. The source code is
annotated using proprietary extensions to the C, C++ or FORTRAN
programming languages that are understood by the \verb+nvcc+
compiler. Recently, CUDA has evolved into an open standard
called~OpenCL that is supported by other vendors; however, in the
following we consider CUDA since is currently more robust and
efficient than OpenCL.

The basic unit of work that can be executed on a CUDA-capable~\ac{GPU}
is the \emph{CUDA thread}. Threads can be arranged in one-, two-, or
three-dimensional \emph{blocks}, that can be further assembled into a
one-, two-, or three-dimensional \emph{grid}. Each thread has unique
identifiers that can be used to map a thread to one input element.
The CUDA paradigm favors decomposition of a problem into very small
tasks that are assigned to threads (\emph{fine-grained
  parallelism}). The~CUDA runtime schedules threads to cores for
execution; the hardware supports multitasking with almost no overhead,
so that the number of threads can (and usually does) exceed the number
of cores.

Implementation of the~\ac{BML} model with CUDA is quite simple, and
consists of transforming the \verb+horizontal_step+ and
\verb+vertical_step+ functions to \emph{CUDA kernels}, i.e., blocks of
code that can be executed by a thread.  CUDA kernels are designated
with the \verb+__global__+ specifier.  Using two-dimensional blocks of
threads it is possible to assign one thread to each element $(i, j)$;
this requires launching $N \times N$ threads. The
\verb+horizontal_step+ function becomes a CUDA kernel as follows:

\begin{lstlisting}
__global__
void horizontal_step(cell_t *cur, cell_t *next, int N) {
    const int i = 1 + threadIdx.y + blockIdx.y * blockDim.y;
    const int j = 1 + threadIdx.x + blockIdx.x * blockDim.x;
    if ( i < N+1 && j < N+1) {
      /* update cell (i,j) */
    }
}
\end{lstlisting}

CUDA cores have no direct access to system RAM; instead, they can only
use the~\ac{GPU} memory, called \emph{device RAM}. Therefore, input
data must be transferred from system~RAM to device~RAM before the CUDA
threads are activated. Once computation on the~\ac{GPU} is completed,
output data is transferred back to system~RAM. The parameters
\verb+cur+ and \verb+next+ above point to device RAM.

\section{Performance evaluation}\label{sec:performance-evaluation}

\begin{table}[t]
\centering%
\caption{Hardware used for the experimental evaluation}\label{tab:hardware}
\begin{tabular}{@{\extracolsep{1.5em}}lll}
\toprule
                & \textbf{Machine~A} & \textbf{Machine~B} \\
\midrule
CPU             & Intel Xeon E3-1220 & Intel Xeon E5-2603 \\
Cores           & 4             & 12                    \\
HyperThreading  & No            & No                    \\
Max CPU freq.   & 3.50~GHz      & 1.70~GHz              \\
L2 Cache        & 256~KB        & 256~KB                \\
L3 Cache        & 8192~KB       & 15360~KB              \\
RAM             & 16~GB         & 64~GB                 \\
\midrule
GPU             & Quadro K620   & GeForce GTX 1070      \\
GPU max clock rate & 1.12~GHz   & 1.80~GHz              \\
Device RAM      & 1993~MB       & 8114~MB               \\
CUDA cores      & 384           & 1920                  \\
\bottomrule
\end{tabular}
\end{table}

In this section we compare the implementations of the~\ac{BML}
automaton described so far: the scalar version \emph{without} ghost
cells from Section~\ref{sec:serial} (\textbf{serial}); the serial
version \emph{with} ghost cells (\textbf{Serial+halo}); the OpenMP
version from Section~\ref{sec:openmp} (\textbf{OpenMP}); the SIMD
version from Section~\ref{sec:simd} (\textbf{SIMD}); a combined
OpenMP+SIMD version, where cells are updated using SIMD instructions
and the outer loops are parallelized with OpenMP directives
(\textbf{OpenMP+SIMD}); finally, the CUDA version from
Section~\ref{sec:cuda} (\textbf{CUDA}). In the cases where~OpenMP is
used, we make use of all processor cores available in the machine.

All implementations have been realized under Ubuntu Linux
version~16.04.4 using GCC~5.4.0 with level~3 optimization enabled
using the compiler flags \verb+-O3 -march=native+; the CUDA version
has been compiled with the proprietary \verb+nvcc+ compiler from the
CUDA Toolkit version~9.1. We run the programs on two multi-core
machines equipped with CUDA capable~\acp{GPU}, whose specifications
are shown in Table~\ref{tab:hardware}. Machine~A has a fast, four-core
processor but includes a low-end~\ac{GPU}. Machine~B has a more
powerful~\ac{GPU} and a processor with twelve cores; however, each
core runs at a lower clock rate.

\begin{figure}[t]
  \centering%
  \includegraphics[scale=.7]{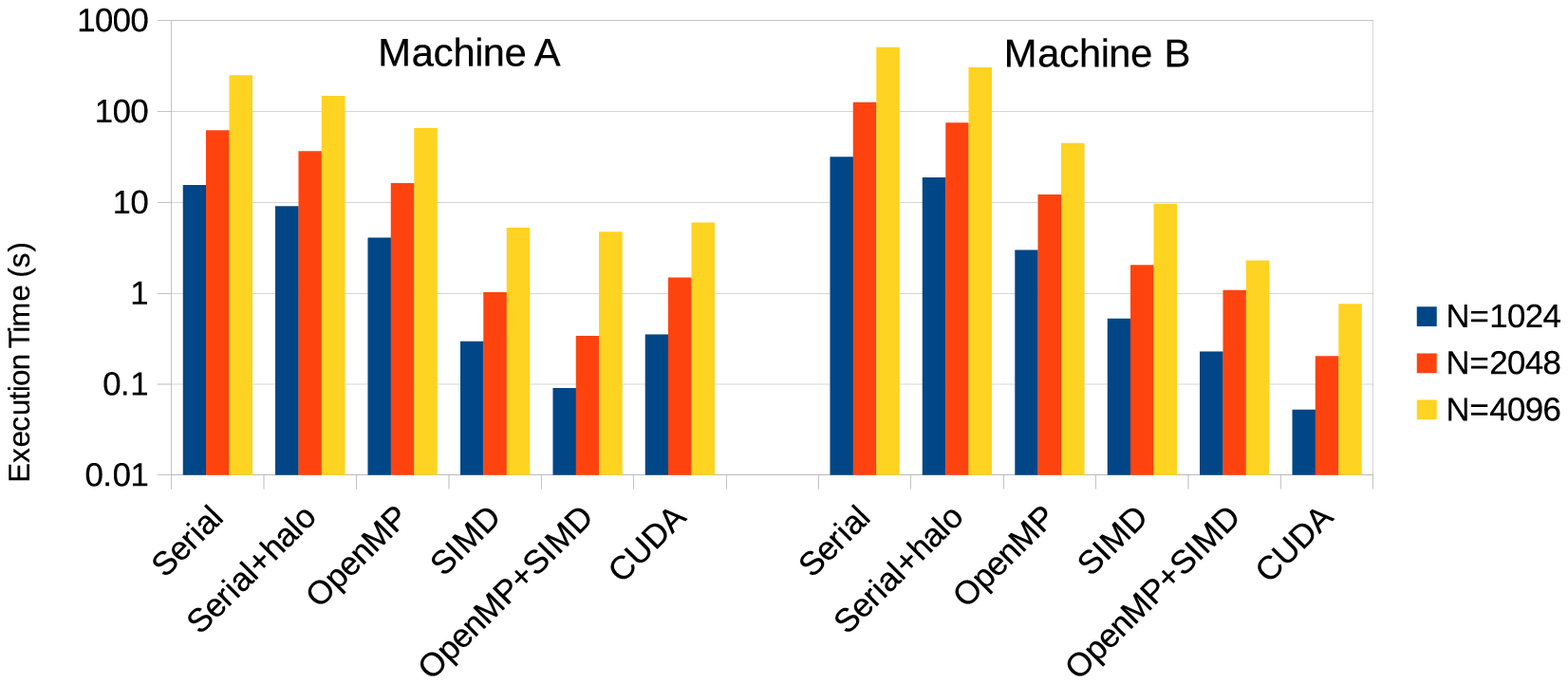}\\
  \begin{tabular}{@{\extracolsep{1em}}lrrrrrr}
    \toprule
    & \multicolumn{3}{c}{Machine~A} & \multicolumn{3}{c}{Machine~B} \\
    \cmidrule{2-4} \cmidrule{5-7}
    $N$         & $1024$ & $2048$ &  $4096$ & $1024$ &  $2048$ &  $4096$ \\
    \midrule
    Serial      & 15.446 & 61.957 & 249.026 & 31.554 & 126.298 & 508.046 \\
    Serial+halo &  9.091 & 36.545 & 148.266 & 18.725 &  75.046 & 304.818 \\
    OpenMP      &  4.080 & 16.256 &  65.366 &  2.988 &  12.184 &  44.763 \\
    SIMD        &  0.294 &  1.021 &   5.252 &  0.524 &   2.044 &   9.652 \\
    OpenMP+SIMD &  0.090 &  0.339 &   4.730 &  0.228 &   1.079 &   2.292 \\
    CUDA        &  0.351 &  1.482 &   5.970 &  0.052 &   0.202 &   0.768 \\
    \bottomrule
  \end{tabular}  
  \caption{Mean execution time of~\ac{BML} model implementations
    ($\rho = 0.3$, $1024$ steps)}\label{fig:execution-times}
\end{figure}

The~\ac{BML} automaton has been simulated with a vehicle density
$\rho=0.3$ on a domain of size $N \times N$, $N=1024, 2048, 4096$
for~$1024$ steps. Figure~\ref{fig:execution-times} shows the execution
time of each program; each measurement is the average of five
execution. Note the logarithmic scale of the vertical axis, which is
necessary since the execution times vary more than two orders of
magnitude.

The use of ghost cells provides a significant reduction of the
execution time (about 40\%) compared to the use of the modulo
operator; note how a simple modification can make such a
difference. The~OpenMP version using all available processor cores
provides an additional speedup of about $2 \times$ for machine~A and
about $6 \times$ for machine~B (recall that machine~B has more
cores). These speedups come very cheaply: the OpenMP version differs
from the serial implementation by a couple of
\verb+#pragma omp parallel for+ directives that have been added to the
functions computing the horizontal and vertical steps.

The SIMD implementation provides perhaps the most surprising
results. On machine~A, a single core delivers more computing power
than the mid-range \ac{GPU} installed; by combining~OpenMP and~SIMD
instructions it is possible to further reduce the execution time.  On
machine~A the OpenMP+SIMD version is more than four times faster than
the~\ac{GPU} for $N=1024, 2048$; however, the gap closes for the
larger domain size $N=4096$. Machine~B has a slower~CPU and a
better~\ac{GPU}, so the~\ac{GPU} version is four to five times faster
than the OpenMP+SIMD version.

\section{Conclusions}\label{sec:conclusions}

In this paper we have analyzed the impact of several implementations
of the~\ac{BML} traffic model on modern CPUs and GPUs. Starting with a
serial version, we have applied the \emph{ghost cells} pattern to
reduced the overhead caused by the access to the neighbors of each
cell. A parallel version has then been derived by applying OpenMP
preprocessor directives to the serial implementation, to take advantage
of multicore processors. More effort is required to restructure the
code to take advantage of~\ac{SIMD} instructions; the payoff is
however surprising: the~\ac{SIMD} version running on a single~CPU core
proved to be faster than a~\ac{GPU} implementation running on a
mid-range graphic card.

The results suggest that traffic models can greatly benefit from
accurately tuned CPU implementations, especially considering that a
fast CPU can execute this type of workload faster than an
average~\ac{GPU}. However, the most useful optimization, namely, the
use of~\ac{SIMD} instructions, requires technical knowledge that the
average user is unlikely to possess. It is therefore advised that
traffic simulators and other \ac{CA} modeling tools make these
features available to the scientific community.

We are extending the work described in this paper by considering more
complex and realistic traffic models based on~\ac{CA}. We expect that
the findings reported above will still apply, at least to a certain
extent, to any discrete~\ac{CA}.

\end{document}